\documentclass[slac_one]{revtex4}

\usepackage{graphicx}

\usepackage{fancyhdr}

\usepackage{wrapfig}
\usepackage{units}
\usepackage{floatflt}

\newcommand{\dEdx}{\ensuremath{\mathrm{d}E/\mathrm{d}x}}
\newcommand{\PYTHIA}{\textsc{Pythia}}
\newcommand{\pslash}{p\llap{/\kern+0.1em}}
\newcommand{\GeV}{\mathrm{G\kern-0.01eme\kern-0.11emV}}
\newcommand{\MeV}{M\kern-0.02eme\kern-0.11emV}
\newcommand{\keV}{ke\kern-0.11emV}
\newcommand{\eV}{e\kern-0.11emV}

\begin{document}


\title{Beauty photoproduction using decays into electrons at HERA} 


%




\author{M. J\"ungst}

\affiliation{Physikalisches\phantom{0}Institut\phantom{0}Universit\"at\phantom{0}Bonn\\
  Nu{\ss}allee\phantom{0}12,\phantom{0}53115\phantom{0}Bonn,\phantom{0}Germany~\hfill}
%



\begin{abstract}

  Photoproduction of beauty quarks in events with two jets and
  an electron associated with one of the jets has been studied with
  the ZEUS detector at HERA using an integrated luminosity of
  $120 \:\text{pb}^{-1}$.  The fractions of events containing $b$ quarks, and also
  of events containing $c$ quarks, were extracted from a likelihood
  fit using variables sensitive to electron identification as well as
  to semileptonic decays.  Total and differential cross sections for
  beauty and charm production were measured and compared with
  next-to-leading-order QCD calculations and Monte Carlo models.
\end{abstract}


\maketitle

\thispagestyle{fancy}





\section{INTRODUCTION}
The production of heavy quarks in $ep$ collisions at HERA is an
important testing ground for perturbative Quantum Chromodynamics
(pQCD) since the large $b$-quark and $c$-quark masses provide a hard
scale that allows perturbative calculations.  When $Q^{2}$, the
negative squared four-momentum exchanged at the electron or
positron is small, the reactions $ep \rightarrow e \, b\bar{b} \, X$ and
$ep \rightarrow e \, c\bar{c} \, X$ can be considered as a photoproduction process in which a
quasi-real photon, emitted by the incoming electron interacts with the
proton. For heavy-quark transverse momenta comparable to the quark mass,
next-to-leading-order (NLO) QCD calculations in which the massive
quark is generated dynamically are expected to provide reliable
predictions for the photoproduction cross sections.
This analysis~\cite{paper} was performed with data taken by the ZEUS~\cite{zeus} detector
from 1996 to 2000, when HERA collided electrons or positrons with energy
$E_e=27.5 \:\text{GeV}$ with protons of energy $E_p=820 \:\text{GeV}$ (1996--1997) or $920 \:\text{GeV}$ (1998--2000).
The corresponding integrated luminosities are $38.6 \pm 0.6 \:\text{pb}^{-1}$ at
centre-of-mass energy $\sqrt{s} = 300 \:\text{GeV}$, and $81.6 \pm 1.8 \:\text{pb}^{-1}$ at
$\sqrt{s} = 318 \:\text{GeV}$.

\section{THEORY}
The measurements are compared to a leading-order plus parton-shower
Monte Carlo (\PYTHIA) as well as QCD predictions at next-to-leading order
(NLO), based on the FMNR programme \cite{frixione}.
This NLO programme separately generates processes containing
point-like and hadron-like photon contributions, which have to be
combined to obtain the total cross section.
The main uncertainties of the NLO calculations originate from the
uncertainties of the heavy-quark masses (pole masses) and the renormalisation
and factorisation scales. The central values for the masses were set to
$m_{b}=4.75 \:\text{GeV}$  and $m_{c}=1.6 \:\text{GeV}$ where both masses were
 varied by $\pm 0.25 \:\text{GeV}$.
The renormalisation, $\mu_{R}$, and factorisation, $\mu_{F}$, scales were chosen to be equal and set to
$\mu_{R}=\mu_{F}=\sqrt{\hat{p}_{T}^{2}+m_{b(c)}^{2}}$, and varied by a factor
two for the uncertainty.

\section{SIGNAL EXTRACTION}
Electron candidates were selected by requiring tracks fitted to
the primary vertex and having a transverse momentum, $p_{T}^{e}$,
of at least $0.9 \:\text{GeV}$ in the pseudorapidity range $|\eta^{e}| < 1.5$.
For the identification of electrons from semileptonic heavy-quark
decays, variables for particle identification were combined with
event-based information characteristic of heavy-quark production.
\vspace{1.0cm}
\subsection{Electron identification}
\begin{wrapfigure}[12]{r}{0.28\textwidth}
  \vspace{-30pt}
  \begin{center}
\includegraphics[bb = 0 0 270 305, clip=true,width=0.23\textwidth]{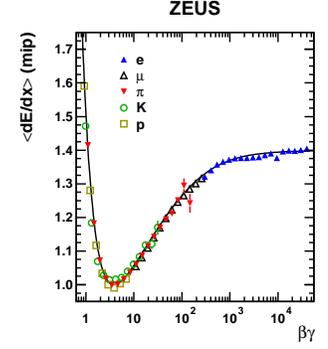}
\end{center}
  \vspace{-20pt}
\caption{The mean \dEdx ~measured in the CTD as a function of $\beta\gamma$.}
\label{dedx}
\end{wrapfigure}
A central tool for this analysis was the \dEdx ~measurement from
the Central Tracking Detector (CTD). The pulse height of the signals on the sense wires was
used as a measure of the specific ionisation.  This pulse
height was corrected for a number of effects ~\cite{detlef}.
After all corrections, the measured \dEdx ~depended only on the
particle velocity, $\beta\gamma$.
This is illustrated in Fig.~\ref{dedx}. It shows the specific energy loss as a
function of $\beta\gamma$, for the different samples of identified particles,
$e^{\pm}, \mu^{\pm}, \pi^{\pm}, K^{\pm}, p, \bar{p}$.
Additional variables for the electron identification are the fraction of
energy in the calorimeter which is deposited in the electromagnetic part of the
calorimeter and the ratio of this energy to the track momentum measured by the
CTD. These two variables use the differences in shower and cluster topologies of
electrons, hadrons and muons.

\subsection{Decay identification}
To identify electrons from semileptonic decays, the event signature of a lepton
from a heavy quark and missing transverse momentum from the neutrino was
used. The size of the transverse-momentum component of the electron
candidate relative to the direction of the jet axis reflects the mass of the
decaying hadron and gives a good separation of the semileptonic $b$-quark
decays from other sources. To distinguish semileptonic $c$-quark as well as
$b$-quark decays from the light flavour background the difference of azimuthal
angles between the electron candidate and the missing transverse momentum vector
was used.

\subsection{Test function}
The discriminating input variables were combined in a likelihood hypothesis
test in order to calculate the heavy quark contributions in the data set.
For a given hypothesis of particle, $i$, and source $j$, the
likelihood, $\mathcal{L}_{ij}$, is given by
\begin{wrapfigure}[13]{r}{0.35\textwidth}
\vspace{-0.5cm}
\includegraphics[width=5.5cm]{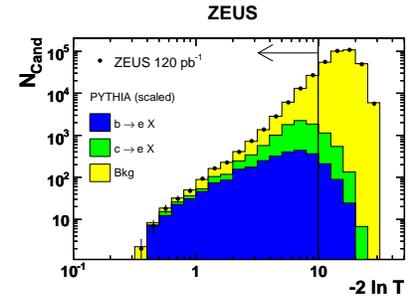}
\caption{The distribution of the likelihood ratio.}
\label{fit}
\end{wrapfigure}
\begin{equation}
  \mathcal{L}_{ij} = \prod\limits_{l}\,\mathcal{P}_{ij}(d_{l})\,,
 \end{equation}
where $\mathcal{P}_{ij}(d_{l})$ is the probability to observe particle
$i$ from source $j$ with value $d_{l}$ of a discriminant variable.
The particle hypotheses $i \in \{e,\mu,\pi,K,p\}$ and sources, $j$,
for electrons from semileptonic beauty, charm decays and background,
$j \in \{b,c,\mathrm{Bkg}\}$, were considered.
For the likelihood ratio test, the test function, $T_{ij}$ was defined as
\begin{equation}
  T_{ij} = \frac{\alpha_{i} \alpha'_{j} \mathcal{L}_{ij}}{%
    \sum \limits_{m,n} \alpha_{m} \alpha'_{n} \mathcal{L}_{mn}}.
\end{equation}
Test functions were calculated separately for the three samples. \\The fractions of the
three samples in the data, $f_{e,b}^{\mathrm{DATA}}$,
$f_{e,c}^{\mathrm{DATA}}$, $f_{\mathrm{Bkg}}^{\mathrm{DATA}}$, were
obtained from a three-component maximum likelihood fit
to the $T$ distributions. The fit (Fig.~\ref{fit}) range of the test function was restricted
to $-2 \ln T < 10$.

\section{RESULTS}
The visible $ep$ cross sections for $b$-quark and
$c$-quark production and the subsequent semileptonic decay to an
electron with $p_{T}^{e} > 0.9 \:\text{GeV}$ in the range $|\eta^{e}| < 1.5$ in
photoproduction events with $Q^{2} < 1 \:\text{GeV}^{2}$ and $0.2 < y < 0.8$ and at
least two jets with $E_{T} > 7 (6) \:\text{GeV}$, $|\eta| < 2.5$ were
determined separately for $\sqrt{s}=300 \:\text{GeV}$ and $\sqrt{s}=318 \:\text{GeV}$.
\begin{wrapfigure}[14]{r}{0.45\textwidth}
  \vspace{-10pt}
  \begin{center}
    \includegraphics[bb = 0 0 480 530, clip=true, width=65mm]{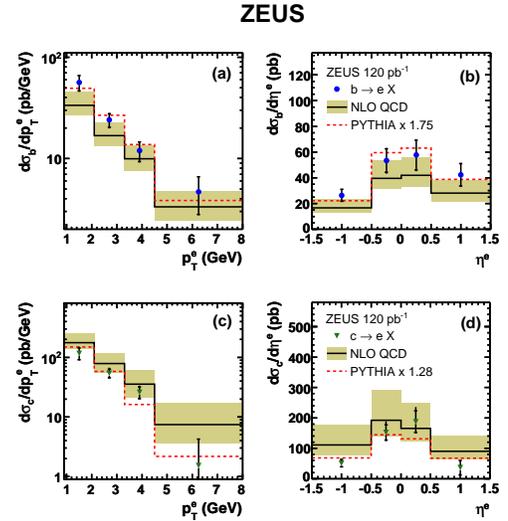}
  \end{center}
  \vspace{-20pt}
  \caption{Differential cross sections  as a function of the a),c) transverse momentum
    and the b),d) pseudorapidity of the electrons for a),b) $b\bar{b}$ and c),d) $c\bar{c}$ production.}\label{diffcross}
\end{wrapfigure}
The cross sections at the two different centre-of-mass energies are
consistent with each other; combining the results leads to a reduced
statistical uncertainty.
For the complete data set ($96-00$) the cross sections and the NLO predictions
evaluated at $\sqrt{s}=318 \:\text{GeV}$ are:\\[0.5cm]
\begin{tabular}{l r l l}
\phantom{aaaaaaa}&$\sigma_{b}^{vis}$ & $= ( 125 \pm 11 (stat.) ^{+10}_{-11} (syst.))$ & pb,\\
&$\sigma_{c}^{vis}$ & $= ( 278 \pm 33 (stat.) ^{+48}_{-24} (syst.))$ & pb.\\[0.5cm]
\end{tabular}
\begin{tabular}{l r l l}
  \phantom{aaaaaaaaaaaaa}&$\sigma_{b}^{NLO}$ & $= 88^{+22}_{-13}$ & pb,\\
                   &$\sigma_{c}^{NLO}$ & $= 380^{+170}_{-110}$ & pb.\\[0.5cm]
\end{tabular}

Differential cross sections as a function of $p_{T}^{e}$ and $\eta_{e}$ are
shown in Fig ~\ref{diffcross}.
The figure also shows the NLO QCD and the scaled \PYTHIA\phantom{0}predictions.
The scale factors have been  calculated from the total visible $b\bar{b}$ and
$c\bar{c}$ cross section, which are a factors of 1.75 and 1.28 higher than
the corresponding \PYTHIA\phantom{0}predictions.
Both the predictions from the NLO QCD calculations as well as the scaled
\PYTHIA\phantom{0}cross sections describe the data well.

\section{CONCLUSIONS}
Beauty and charm production have been measured in dijet photoproduction using
semileptonic decays into electrons.
The results were compared to both NLO QCD calculations as well as
predictions from a Monte Carlo model. The NLO QCD predictions are consistent with the data.

\begin{wrapfigure}[11]{r}{0.45\textwidth}
  \vspace{-35pt}
  \begin{center}
\includegraphics[width=75mm]{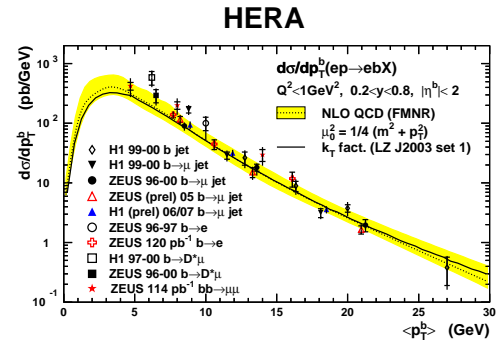}
  \end{center}
  \vspace{-20pt}
\caption{Cross sections for beauty production as a function of $p^{b}_{T}$
for various decay channels.} \label{beauty}
\end{wrapfigure}
 The Monte Carlo models describe well the shape of the differential distributions in the data.
The good agreement with the NLO QCD prediction allows the cross section as a function of $p_{T}^{b}$ to be
extracted. The resulting cross section is shown in Fig ~\ref{beauty} and is
also compared with previous measurements by both the H1 and ZEUS collaborations.
The measurements agree well with the previous values, giving a consistent picture of
$b$-quark production in $ep$ collisions in the photoproduction regime,
and are well reproduced by the NLO QCD calculations.

\end{document}